\documentclass[letterpaper]{ptephy}

\newcommand{{\SlashD}}{D\!\!\!\!\big/}

\begin{document}

\title{Tera Scale Remnants of Unification and Supersymmetry
at Planck Scale}

\author{\name{\fname{Yoshiharu} \surname{Kawamura}}{\ast}
}

\address{\affil{}{Department of Physics, Faculty of Science, Shinshu University,
Matsumoto 390-8621, Japan}
${}^\ast$\!\! {\rm E-mail: haru@azusa.shinshu-u.ac.jp}
}

\begin{abstract}%
We predict new particles at the Tera scale
based on the assumptions that the standard model gauge interactions are unified 
around the gravitational scale with a big desert
and new particles originate from hypermultiplets as remnants of supersymmetry,
and propose a theoretical framework at the Tera scale and beyond,
that has predictability.
\end{abstract}

\subjectindex{B40, B42}

\maketitle

\section{Introduction}\label{Introduction}

Various experimental results have provided useful hints to explore the physics beyond the standard model (SM).

The first hint is the precision measurements of gauge coupling constants at LEP \cite{LEP}.
Those values suggest that the gauge interactions can be unified at a high-energy scale $M_{\tiny{\mbox{U}}}$,
with the advent of new particles \cite{U1,U2,U3,U4}.

The second one is the discovery of Higgs boson at LHC \cite{LHC1,LHC2}.
{}From the observed value of mass $m_h \doteqdot 126$GeV, 
the quartic self-coupling constant of Higgs boson
is estimated as $\lambda \doteqdot 0.131$, using the mass formula $m_h = \sqrt{2\lambda}v$
and the vacuum expectation value (VEV) of Higgs boson $v=246$GeV.

The third one is that superpartners have not yet been discovered.
The gauge hierarchy problem revisits, 
in case that the physics beyond the Tera scale is relevant to the SM \cite{GH1,GH2}.
If we require that $m_h$ should be determined without a fine tuning,
we have an option that there is a big desert from the Tera scale to $M_{\tiny{\mbox{U}}}$,
and the initial value of $m_h$ is fixed by the physics at $M_{\tiny{\mbox{U}}}$  
(protected against large radiative corrections by some mechanism or symmetry) \cite{A&I}.
Hence new charged particles must appear around the Tera scale if exist until $M_{\tiny{\mbox{U}}}$.

The last one is that a definite discrepancy has not yet been observed 
between the predictions in the SM (modified with massive neutrinos) 
and experimental results.
This suggests that, in case with new particles, 
their effects must be canceled or suppressed for some reason.
A clue might be a supersymmetry (SUSY), which can induce a cancellation 
between contributions from bosons and those from fermions 
and control interactions among particles \cite{SUSY}.
There is a possibility that new massive particles 
originate from hypermultiplets
and a power of SUSY remains partially and indirectly, 
although SUSY is broken down around $M_{\tiny{\mbox{U}}}$.

Therefore it is interesting to pursue the physics at the Tera and unification scale
based on the above hints.
In this letter, we specify new particles at the Tera scale
based on the assumptions that the SM gauge interactions are unified 
around the gravitational scale with a big desert
and new particles come from hypermultiplets.
We propose a theory at the Tera scale and beyond, 
that has a power to determine some coupling constants.

The outline of this letter is as follows.
In the next section, we explore new particles at the Tera scale, as remnants of
gauge unification and SUSY.
We construct a theory including new particles in section 3.
In the last section, we give conclusions and discussions.

\section{New particles at Tera scale}
\label{New particles at Tera scale}

We study new particles at the Tera scale based on the following assumptions.

One is related to the grand unification.
We assume that the values of gauge coupling constants $g_i$ $(i=1,2,3)$ agree 
at the gravitational scale $M \equiv M_{\tiny{\mbox{Pl}}}/\sqrt{8\pi} = 2.4 \times 10^{18}$GeV
($M_{\tiny{\mbox{Pl}}}$ is the Planck mass),
and the gauge interactions are unified into a simple group.
Then, the normalization of hypercharge is fixed and the gauge coupling constant relating $U(1)_Y$
is given by $g_1 = \sqrt{5/3} g'$, 
where $g'$ is the gauge coupling constant of $U(1)_Y$ in the SM.
The value of unified one $g_{\tiny{\mbox{U}}}$ is supposed to be less than one, i.e.,
the theory is described as a weakly coupled system at $M$, for simplicity.

The other is related to new particles and SUSY.
We assume that there exist extra particles with a same gauge quantum numbers of SM fermions
and/or its mirror particles around the Tera scale,
there is a big desert beyond the Tera scale until $M$, 
except for right-handed neutrinos ($\nu_R$),
and the new colored particles form hypermultiplets
as remnants of SUSY, that is broken down at $M$. 

The solutions of renormalization group equations (RGEs) for $g_i$ are expressed as
\begin{eqnarray}
\alpha_i^{-1}(\mu) = \alpha_{\tiny{\mbox{U}}}^{-1} + \frac{b_i}{2\pi} \ln \frac{M}{\mu}~,
\label{RGeq-gi}
\end{eqnarray} 
where $\alpha_i \equiv g_i^2/(4\pi)$,  $\mu$ is a renormalization point
(an arbitrary scale in the big desert),
$b_i$s are coefficients of $\beta$ functions at the one-loop level.
In (\ref{RGeq-gi}), we replace $\alpha_i^{-1}(M)$ with $\alpha_{\tiny{\mbox{U}}}^{-1}$, 
using the unification conditions $\alpha_{\tiny{\mbox{U}}} \equiv g_{\tiny{\mbox{U}}}^2/(4\pi) = \alpha_i(M)$.
Hereafter, we take the $Z^0$ gauge boson mass 
$M_Z(\doteqdot 91.19 \mbox{GeV})$ as $\mu$, for simplicity.

By eliminating $\alpha_{\tiny{\mbox{U}}}^{-1}$,
the unification conditions are written as
\begin{eqnarray}
b_i - b_j = \frac{2\pi [\alpha_i^{-1}(M_Z) - \alpha_j^{-1}(M_Z)]}{\ln ({M}/{M_Z})}~,~~ (i,j=1,2,3)~.
\label{bi-bj}
\end{eqnarray}
By using the experimental values
\begin{eqnarray}
&~& \alpha_1^{-1}(M_Z) \doteqdot 59.01~,~~\alpha_2^{-1}(M_Z) \doteqdot 29.57~,~~
\alpha_3^{-1}(M_Z) \doteqdot 8.446~,
\label{alpha-exp}
\end{eqnarray}
we obtain the relations
\begin{eqnarray}
b_2 - b_3 \doteqdot 3.51~,~~b_1 - b_2 \doteqdot 4.89~,~~b_1 - b_3 \doteqdot 8.40~.
\label{b2-b3}
\end{eqnarray}

In the presence of new particles,
$b_i$s are given by
\begin{eqnarray}
&~& b_1 = \frac{4}{3} N_f + \frac{1}{10} N_h + \frac{1}{30} n_Q + \frac{4}{15} n_U + \frac{1}{15} n_D 
+ \frac{1}{10} n_L + \frac{1}{5} n_E~,
\label{b1}\\
&~& b_2 = -\frac{22}{3} + \frac{4}{3} N_f + \frac{1}{6} N_h + \frac{1}{2} n_Q + \frac{1}{6} n_L~,
\label{b2}\\
&~& b_3 = -11 + \frac{4}{3} N_f + \frac{1}{3} n_Q + \frac{1}{6} n_U + \frac{1}{6} n_D~,
\label{b3}
\end{eqnarray}
where $N_f$ is a number of family, $N_h$ is a number of Higgs doublet, and
$n_Q$, $n_U$, $n_D$, $n_L$ and $n_E$ are numbers of 
fields (counting that of a complex scalar field as one)
with representations $({\bf 3}, {\bf 2}, 1/6)$ or $(\overline{\bf 3}, {\bf 2}, -1/6)$,
$(\overline{\bf 3}, {\bf 1}, -2/3)$ or $({\bf 3}, {\bf 1}, 2/3)$,
$(\overline{\bf 3}, {\bf 1}, 1/3)$ or $({\bf 3}, {\bf 1}, -1/3)$,
$({\bf 1}, {\bf 2}, -1/2)$ or $({\bf 1}, {\bf 2}, 1/2)$
and $({\bf 1}, {\bf 1}, 1)$ or $({\bf 1}, {\bf 1}, -1)$
under $(SU(3)_C, SU(2)_L, U(1)_Y)$, respectively.
The number of a Weyl fermion is considered as double that of a complex scalar field,
and then that of a chiral supermultiplet is considered as triple that of a complex scalar field.
Hence, $n_Q$, $n_U$ and $n_D$ should be multiples of 6,
if new colored particles form hypermultiplets.

Using (\ref{b2-b3}) -- (\ref{b3}),
we derive the relations
\begin{eqnarray}
&~& -n_Q + n_U + n_D - n_L \doteqdot 1.94 ~(\simeq 2)~,
\label{r1}\\
&~& -7 n_Q + 4 n_U + n_D - n_L + 3 n_E \doteqdot - 35.65 ~(\simeq -34)~,
\label{r2}\\
&~& 3 n_Q - n_U + n_D - n_L - 2 n_E \doteqdot 27.0~(\simeq 26)~,
\label{r3}
\end{eqnarray}
where we take $N_f=3$ and $N_h=1$ according to the SM,
and two of (\ref{r1}) -- (\ref{r3}) are independent.

If we impose the condition 
$\alpha_{\tiny{\mbox{U}}} \lesssim 1/10$
on $\alpha_{\tiny{\mbox{U}}}$ based on the assumption that $g_{\tiny{\mbox{U}}}$ is less than one,
we obtain the inequalities
\begin{eqnarray}
n_Q + 8 n_U + 2 n_D + 3 n_L + 6 n_E \lesssim 121~,~~
3 n_Q +  n_L \lesssim 38.5~,~~
2 n_Q + n_U + n_D \lesssim 40.0~.
\label{ineq}
\end{eqnarray}

Using (\ref{r1}) -- (\ref{ineq}),
we obtain {\it a unique solution}
\begin{eqnarray}
n_Q = 6~,~~ n_U = 0~,~~ n_D = 12~,~~ n_L = 4~,~~ n_E = 0~,
\label{sol}
\end{eqnarray}
where we use values in parenthesis in (\ref{r1}) -- (\ref{r3}), for the sake of convenience,
and require that $n_Q$, $n_U$ and $n_D$ should be multiples of 6.
The value of unified gauge coupling constant is estimated as 
$\alpha_{\tiny{\mbox{U}}} = \alpha_i(M) \sim {1}/{26}$.

The uniqueness of (\ref{sol}) depends on our assumptions, that are backed by not so strong evidence.
Another solutions are obtained by changing assumptions partially.
Here, we give typical examples.\\
(1) Change of unification scale :
If we take $M_{\tiny{\mbox{U}}} \simeq 10^{17}$GeV in place of $M$,
we obtain a solution $n_Q = 6$, $n_U = 0$, $n_D = 6$, $n_L = 0$ and $n_E = 2$
with $\alpha_{\tiny{\mbox{U}}} \sim 1/31$.\\
(2) Change of normalization of $U(1)_Y$ :
If we take $g_1 = \sqrt{1.1} g'$ in place of $g_1 = \sqrt{5/3} g'$,
we obtain a solution $n_Q = 0$, $n_U = 0$, $n_D = 6$, $n_L = 4$ and $n_E = 0$
with $\alpha_{\tiny{\mbox{U}}} \sim 1/45$ and $M_{\tiny{\mbox{U}}} = M$.
On unification based on string models, the change of normalization 
is related to the choice of Kac-Moody level \cite{String}.\\
(3) Beyond weakly coupled region :
There is a solution $n_Q = 12$, $n_U = 12$, $n_D = 6$, $n_L = 4$ and $n_E = 0$
(or $n_Q = 12$, $n_U = 6$, $n_D = 12$, $n_L = 4$ and $n_E = 6$)
with $\alpha_{\tiny{\mbox{U}}} \sim 1/8.4$ and $M_{\tiny{\mbox{U}}} = M$.

Furthermore, a variety of solutions are obtained if we relax the assumption that new colored 
particles form hypermultiplets.
For example, there is a solution $n_Q = 6$, $n_U = 0$, $n_D = 8$, $n_L = 0$ and $n_E = 0$
with $\alpha_{\tiny{\mbox{U}}} \sim 1/31$ and $M_{\tiny{\mbox{U}}} = M$.

\section{Theory at Tera scale and beyond}
\label{Theory at Tera scale and beyond}

We construct a theory including new particles
without specifying its particle contents, to maintain a generality.

The Lagrangian density of theory in a big desert is given by
\begin{eqnarray}
&~& \mathcal{L}_{\tiny{\mbox{BSM}}} = \mathcal{L}_{\tiny{\mbox{SM}}} 
+ \sum_{k} (|D_{\mu} \phi^k|^2 - M_{k}^2 |\phi^k|^2)
+ \sum_{d} \overline{\psi}^d (i \SlashD + M_d) \psi^d
\nonumber \\
&~& ~~~
- \sum_k \lambda_k |\phi^k|^4 
- \sum_{k < l} \lambda_{kl} |\phi^k|^2 |\phi^l|^2
- \sum_{k} \lambda'_{k} |\phi^k|^2 |H|^2 + \mathcal{L}_{\tiny{\mbox{Ynew}}}~,
\label{L}
\end{eqnarray}
where $\mathcal{L}_{\tiny{\mbox{SM}}}$ is the SM one and
new particles are denoted by $\phi^k$ and $\psi^d$.
Here, $\phi^k$s represent complex scalar fields,
$\psi^d$s represent Dirac fermions $\psi^d=(\psi_L^d, \psi_R^d)$,
$H$ is the Higgs doublet of SM, $|\phi^k|^4=(|\phi^k|^2)^2$, 
and $\mathcal{L}_{\tiny{\mbox{Ynew}}}$ stands for Yukawa couplings
such as $y'_{de} \overline{\psi}_L^d H \psi_R^e$ with a suitable assignment of the SM gauge quantum numbers.

Here, we present a conjecture that {\it the theory described by (\ref{L}) 
is derived from a theory at $M_{\tiny{\mbox{U}}}$, 
after the breakdown of a unified gauge symmetry and SUSY, 
and a possible candidate is a unified theory with $N=2$ SUSY 
as an effective theory in the unbroken phase.}

First, we give a simple model to illustrate our idea.
Let the relevant part of $\mathcal{L}_{\tiny{\mbox{BSM}}}$ originate from the following
Lagrangian density given in the unbroken phase,
\begin{eqnarray}
&~& \mathcal{L}_{\tiny{\mbox{new}}}^{N=2} = 
\sum_{\tiny{\mbox{R}}} (|D_{\mu} \Phi_{\tiny{\mbox{R}}}|^2 - M_{\tiny{\mbox{R}}}^2 |\Phi_{\tiny{\mbox{R}}}|^2)
+ \sum_{\tiny{\mbox{R}}} \overline{\Psi}_{\tiny{\mbox{R}}} (i \SlashD + M_{\tiny{\mbox{R}}}) \Psi_{\tiny{\mbox{R}}}
\nonumber \\
&~& ~~~ 
- \frac{1}{2} g_{\tiny{\mbox{U}}}^2 \sum_A \left(\sum_{\tiny{\mbox{R}}} \Phi^{\dagger}_{\tiny{\mbox{R}}} T^A 
  \Phi_{\tiny{\mbox{R}}} 
+ if^{ABC} \overline{\Phi}_{\tiny{\mbox{Adj}}}^B \Phi_{\tiny{\mbox{Adj}}}^C\right)^2
- 2 g_{\tiny{\mbox{U}}}^2 \sum_A \left|\sum_{\tiny{\mbox{R}}} \Phi_{\overline{\tiny{\mbox{R}}}} T^A 
  \Phi_{\tiny{\mbox{R}}}\right|^2~,
\label{LnewN=2}
\end{eqnarray}
where $\Phi_{\tiny{\mbox{R}}}$ and $\Psi_{\tiny{\mbox{R}}}$ form 
a hypermultiplet with a representation $\mbox{R}$ of a unified group $G_{\tiny{\mbox{U}}}$
and $\Phi_{\tiny{\mbox{Adj}}}^A$s are scalar fields that are members of gauge supermultiplets.
The first term in the second line of (\ref{LnewN=2}) comes from $D$-term,
while the second one comes from $F$-term relating $\Phi_{\tiny{\mbox{Adj}}}^A$.
Note that the term such as $\Phi_{\overline{\tiny{\mbox{R}}}} \Phi_{\tiny{\mbox{Adj}}} \Phi_{\tiny{\mbox{R}}}$
is gauge invariant and allowed in the superpotential, 
where $\overline{\mbox{R}}$ is the complex conjugate representation of $\mbox{R}$.

Unless sizable contributions appear on the breakdown of symmetries,
the theory has predictability for some coupling constants,\footnote{
Other type of unified theories called Finite Unified Theories, which have a large predictive power,
have been proposed \cite{FUT}.
They are based on finiteness and the principle of reduction of coupling constants.
}
and our conjecture can be tested by studying the flow of various coupling constants under RGEs,
where the initial values are fixed as
\begin{eqnarray}
&~& \alpha_i(M_{\tiny{\mbox{U}}}) = \alpha_{\tiny{\mbox{U}}}~,~~
\lambda(M_{\tiny{\mbox{U}}}) = c_{\lambda} \alpha_{\tiny{\mbox{U}}}~,~~
\lambda_k(M_{\tiny{\mbox{U}}}) = c_k \alpha_{\tiny{\mbox{U}}}~,~~
\lambda_{kl}(M_{\tiny{\mbox{U}}}) = c_{kl} \alpha_{\tiny{\mbox{U}}}~,
\nonumber \\
&~& \lambda'_{k}(M_{\tiny{\mbox{U}}}) = c'_{k} \alpha_{\tiny{\mbox{U}}}~,~~
y'_f(M_{\tiny{\mbox{U}}}) = 0~,
\label{lambdaM}
\end{eqnarray}
where $c_{\lambda}$, $c_k$, $c_{kl}$ and $c'_k$ are model-dependent constants, e.g.,
$c_{\lambda}=8\pi/25$ and $c_k=24\pi/25$ 
for $G_{\tiny{\mbox{U}}}=SU(5)$,
and $H$ and $\phi^k$ belong to 5-plet and 10-plet, respectively.
The $y'_f$s are Yukawa coupling constants including $\mathcal{L}_{\tiny{\mbox{Ynew}}}$.
Note that any Yukawa couplings in $\mathcal{L}_{\tiny{\mbox{Ynew}}}$ are not allowed at $M_{\tiny{\mbox{U}}}$, 
though Yukawa couplings including gauginos exist in the unbroken phase.
The fixation of coupling constants is due to the fact that
the only interactions of hypermultiplets are gauge interactions if $N=2$ SUSY is respected.

The RGE of $\lambda$ (the quartic coupling constant of $H$) at the one loop level is given by
\begin{eqnarray}
&~& \frac{d \lambda}{d t} = 24 \lambda^2 - 6 y_t^4 + 12 \lambda y_t^2 - 3 \lambda (g'^2 + 3 g^2)
+ \frac{3}{8}\left[2g^4 + (g'^2 + g^2)^2\right] + T_{\tiny{\mbox{new}}}~,
\label{lambda}\\
&~& T_{\tiny{\mbox{new}}} = \sum_k a_k {\lambda'_k}^2 - \sum_f b_f {y'_f}^4
+ \sum_{k, f} b_{kf} {\lambda'_k} {y'_f}^2~,
\label{Tnew}
\end{eqnarray}
where $t \equiv (1/16\pi^2)\ln (\mu/M_{\tiny{\mbox{U}}})$, $y_t$ is Yukawa coupling constant of the top quark, 
$g$ is the gauge coupling constant of $SU(2)_L$,
$T_{\tiny{\mbox{new}}}$ stands for contributions from new particles, and
$a_k$, $b_f$ and $b_{kf}$ are model-dependent constants.
It is interesting to find a model to derive $\lambda \doteqdot 0.131$.

If new particles couple to the SM fermions and the Higgs boson very weakly
at the Tera scale,
corrections to the SM predictions could be negligibly small
and it could explain the agreement between the theoretical values and experimental ones
based on the SM.

At this stage, following questions are left unanswered.
What is the origin of weak scale, i.e. $v=246$GeV?
What is the origin of new particles' masses at the Tera scale?

These can be explained by an extension with an extra $U(1)$ gauge symmetry 
($U(1)_{\tiny{\mbox{C}}}$) and an SM singlet complex scalar field $S$.
In this case, the following terms are added to $\mathcal{L}_{\tiny{\mbox{BSM}}}$,
\begin{eqnarray}
\mathcal{L}_{\tiny{\mbox{newS}}} = 
|D_{\mu} S|^2 - M_{S}^2 |S|^2
- \sum_k \lambda_{Sk} |S|^2 |\phi^k|^2 - \lambda_{SH} |S|^2 |H|^2 - \lambda_{S} |S|^4~.
\label{LnewS}
\end{eqnarray}

By requiring that all hypermultiplets are massless at $M_{\tiny{\mbox{U}}}$,
(\ref{LnewN=2}) is replaced by
\begin{eqnarray}
\hspace{-1.4cm}&~& \mathcal{L}_{\tiny{\mbox{newC}}}^{N=2} = 
\sum_{\tiny{\mbox{R}}} |D_{\mu} \Phi_{\tiny{\mbox{R}}}|^2 
+ \sum_{\tiny{\mbox{R}}} \overline{\Psi}_{\tiny{\mbox{R}}} i \SlashD \Psi_{\tiny{\mbox{R}}}
- \frac{1}{2} g_{\tiny{\mbox{U}}}^2 \sum_A \left(\sum_{\tiny{\mbox{R}}} \Phi^{\dagger}_{\tiny{\mbox{R}}} T^A 
  \Phi_{\tiny{\mbox{R}}} 
+ if^{ABC} \overline{\Phi}_{\tiny{\mbox{Adj}}}^B \Phi_{\tiny{\mbox{Adj}}}^C\right)^2
\nonumber \\
\hspace{-1.4cm}&~& ~~~ 
- 2 g_{\tiny{\mbox{U}}}^2 \sum_A \left|\sum_{\tiny{\mbox{R}}} \Phi_{\overline{\tiny{\mbox{R}}}} T^A 
  \Phi_{\tiny{\mbox{R}}}\right|^2
- \frac{1}{2} g_{\tiny{\mbox{C}}}^2 \left(\sum_{\tiny{\mbox{R}}} q_{\tiny{\mbox{R}}}
  |\Phi_{\tiny{\mbox{R}}}|^2 + q_S |S|^2\right)^2
- 2 g_{\tiny{\mbox{C}}}^2 \left|\sum_{\tiny{\mbox{R}}} q_{\tiny{\mbox{R}}} \Phi_{\overline{\tiny{\mbox{R}}}}
  \Phi_{\tiny{\mbox{R}}}\right|^2~,
\label{LnewCN=2}
\end{eqnarray}
where $g_{\tiny{\mbox{C}}}$ is the gauge coupling constant of $U(1)_{\tiny{\mbox{C}}}$,
and $q_{\tiny{\mbox{R}}}$ and $q_S$ are the $U(1)_{\tiny{\mbox{C}}}$ charges of
$\Phi_{\tiny{\mbox{R}}}$ and $S$, respectively.
The $\mathcal{L}_{\tiny{\mbox{newC}}}^{N=2}$ has a classical conformal invariance,
and Bardeen's argument \cite{Bardeen} of radiative corrections on masses
can be applied to our theory.
Strictly speaking, $\mathcal{L}_{\tiny{\mbox{newC}}}^{N=2}$ should be replaced by that in the broken phase.
Our theory can be considered as effective one relating massless states of string,
and we assume that the breakdown of relevant symmetries is due to stringy effects
and the masslessness of particles in $\mathcal{L}_{\tiny{\mbox{BSM}}}$ 
is protected, even in the broken phase, 
by a magic of string or finiteness \cite{String2}.\footnote{
As the conformal invariance is explicitely broken 
in the presence of massive particles at $M_{\tiny{\mbox{U}}}$,
the breakdown of $U(1)_{\tiny{\mbox{C}}}$ 
through the non-minimal coupling between $S$ and gravity \cite{Oda}
does not occur.
}

{}From the matching condition between (\ref{L}), (\ref{LnewS}) and (\ref{LnewCN=2}),
we obtain the relations
\begin{eqnarray}
&~& \alpha_i(M_{\tiny{\mbox{U}}}) = \alpha_{\tiny{\mbox{U}}}~,~~
\lambda(M_{\tiny{\mbox{U}}}) = c_{\lambda} \alpha_{\tiny{\mbox{U}}} + 2\pi q_H^2 \alpha_{\tiny{\mbox{C}}}~,~~
\lambda_k(M_{\tiny{\mbox{U}}}) = c_k \alpha_{\tiny{\mbox{U}}} + 2\pi q_k^2 \alpha_{\tiny{\mbox{C}}}~,~~
\nonumber \\
&~& \lambda_{kl}(M_{\tiny{\mbox{U}}}) = c_{kl} \alpha_{\tiny{\mbox{U}}} + 2\pi q_k q_l \alpha_{\tiny{\mbox{C}}}~,~~
\lambda'_{k}(M_{\tiny{\mbox{U}}}) = c'_{k} \alpha_{\tiny{\mbox{U}}} + 2\pi q_H q_k \alpha_{\tiny{\mbox{C}}}~,~~
y'_f(M_{\tiny{\mbox{U}}}) = 0~,
\nonumber \\
&~& \lambda_{SH}(M_{\tiny{\mbox{U}}}) = 2\pi q_S q_H \alpha_{\tiny{\mbox{C}}}~,~~
\lambda_{Sk}(M_{\tiny{\mbox{U}}}) = 2\pi q_S q_k \alpha_{\tiny{\mbox{C}}}~,~~
\lambda_{S}(M_{\tiny{\mbox{U}}}) = 2\pi q_S^2 \alpha_{\tiny{\mbox{C}}}~,
\label{lambdaMS}
\end{eqnarray}
where $\alpha_{\tiny{\mbox{C}}} \equiv g_{\tiny{\mbox{C}}}^2/(4\pi)$
and the contribution from the last term in (\ref{LnewCN=2}) is omitted, for simplicity.

In the SM, the electroweak symmetry is broken down with $m^2 < 0$ 
for the Higgs potential
\begin{eqnarray}
V = m^2 |H|^2 + \lambda |H|^4~,
\label{V}
\end{eqnarray}
where $v = \sqrt{-m^2/2\lambda}$.
In case with $m^2(M_{\tiny{\mbox{U}}}) =0$ and $M_k^2(M_{\tiny{\mbox{U}}})=0$, 
we obtain the relations
\begin{eqnarray}
m^2 = \lambda_{SH} |\langle S \rangle|^2~~~~(\mbox{or}~~v = \sqrt{-\lambda_{SH}/2\lambda}|\langle S \rangle|)~,~~
M_k^2 = \lambda_{Sk} |\langle S \rangle|^2~,
\label{mMk}
\end{eqnarray}
where $\langle S \rangle$ is the VEV of $S$.
Then, in order to derive $v=246$GeV and $M_k = O(1)$TeV,
we need the conditions
\begin{eqnarray}
\lambda_{SH} < 0~,~~ \lambda_{Sk} > 0~~,~~
|\lambda_{SH}| = O(1/10^2)|\lambda_{Sk}|~,~~ |\langle S \rangle| \ne 0~.
\label{EW}
\end{eqnarray}

A possible scenario to realize (\ref{EW}) is as follows.\footnote{
A basic idea is same as that in \cite{BL1,BL2}, where $U(1)_{B-L}$ plays a role of $U(1)_{\tiny{\mbox{C}}}$.
}
After coupling constants are running from $M_{\tiny{\mbox{U}}}$ 
with initial values 
$\lambda_{SH}(M_{\tiny{\mbox{U}}}) \ge 0$, $\lambda_{Sk}(M_{\tiny{\mbox{U}}}) > 0$ 
and $\lambda_{S}(M_{\tiny{\mbox{U}}}) > 0$, 
$S$ obtains a VEV via the Coleman-Weinberg mechanism \cite{CW},
and the $U(1)_{\tiny{\mbox{C}}}$ is broken down around the Tera scale.
Then, $H$ and $\phi^k$ acquire the mass squared $m^2 = \lambda_{SH} |\langle S \rangle|^2$
and $M_k^2 = \lambda_{Sk} |\langle S \rangle|^2$, respectively.
If $\lambda_{SH}$ takes a negative value with a suitable magnitude around the weak scale,
$SU(2)_L \times U(1)_Y$ can be broken down to $U(1)_{\tiny{\mbox{EM}}}$.
Furthermore if $\lambda_{Sk}$s take positive values with suitable magnitudes,
$\phi^k$s have masses of $O(1)$TeV.
The positivity of $M_k^2$ requires $q_S q_k > 0$,
and it means that $\phi^k$s possess $U(1)_{\tiny{\mbox{C}}}$ charge of the same sign as $q_S$.
In other words, scalar fields with the opposite sign as $q_S$ 
should be decoupled at $M_{\tiny{\mbox{U}}}$, and then hypermultiplets do not survive completely.
It is also interesting to find a model to derive $v = 246$GeV and $M_k = O(1)$TeV.

\section{Conclusions}
\label{Conclusions}

We have predicted new particles at the Tera scale
based on the assumptions that the standard model gauge interactions are unified 
around the gravitational scale with a big desert
and new particles originate from hypermultiplets as remnants of supersymmetry,
and have proposed a theoretical framework at the Tera scale and beyond, 
that has predictability.
The candidate of effective ultra-violet theory is a unified theory 
with $N=2$ SUSY and conformal invariance.
To verify our framework, it is necessary to carry out model-dependent analysis
and to study its particle spectrum and phenomenology. 

The following riddles remain unsolved.
What is the origin of SM fermions?
What is the origin of the hierarchical structure of Yukawa coupling constants?

Chiral superfields based on $N=1$ SUSY might play an important role.
Because a higher-dimensional space-time offers the environment
that both $N=2$ and $N=1$ SUSY coexist,
there is a prospect
that SM particles and new particles originate from 
SUSY orbifold grand unified theories \cite{OGUT1,OGUT2}.

Hence, if predicted particles were discovered at the near future,
there is a possibility that it suggests 
the reality of three big paradigms in particle physics,
i.e., the grand unification, SUSY and extra dimensions.

\ack

The author thanks N. Haba, R. Takahashi, K. Kaneta and I. Oda for valuable discussions.
This work was supported in part by scientific grants 
from the Ministry of Education, Culture,
Sports, Science and Technology under Grant Nos.~22540272 and 21244036 (Y.K.).


\end{document}